\documentclass[conference]{IEEEtran}
\IEEEoverridecommandlockouts
\usepackage{cite}
\usepackage{mdwlist}
\usepackage{amsmath,amssymb,amsfonts,dsfont,booktabs}
\usepackage{algorithm}
\usepackage[noend]{algpseudocode}
\usepackage{graphicx}
\usepackage{subcaption}
\usepackage{textcomp}
\usepackage{xcolor}
\usepackage{float}
\usepackage{authblk}
\usepackage{lipsum}
\def\BibTeX{{\rm B\kern-.05em{\sc i\kern-.025em b}\kern-.08em
    T\kern-.1667em\lower.7ex\hbox{E}\kern-.125emX}}
    
\makeatletter
\def\BState{\State\hskip-\ALG@thistlm}
\makeatother

\begin{document}
\setlength{\abovedisplayskip}{3pt}
\setlength{\belowdisplayskip}{3pt}

\title{DeVLearn: A Deep Visual Learning Framework for Localizing Temporary Faults in Power Systems}
\author{Shuchismita Biswas, Rounak Meyur and Virgilio Centeno
\\Department of Electrical and Computer Engineering, Virginia Tech, Blacksburg, VA, USA\\
\{suchi, rounakm8, virgilio\}@vt.edu}

\renewcommand\Authands{ and }

\maketitle

\begin{abstract}
Frequently recurring transient faults in a transmission network may be indicative of impending permanent failures. Hence, determining their location is a critical task.  
This paper proposes a novel image embedding aided deep learning framework called DeVLearn for faulted line location using PMU measurements at generator buses. Inspired by breakthroughs in computer vision, DeVLearn represents measurements (one-dimensional time series data) as two-dimensional unthresholded Recurrent Plot (RP) images. These RP images preserve the temporal relationships present in the original time series and are used to train a deep Variational Auto-Encoder (VAE). The VAE learns the distribution of latent features in the images. Our results show that for faults on two different lines in the IEEE 68-bus network, DeVLearn is able to project PMU measurements into a two-dimensional space such that data for faults at different locations separate into well-defined clusters. This compressed representation may then be used with off-the-shelf classifiers for determining fault location. The efficacy of the proposed framework is demonstrated using local voltage magnitude measurements at two generator buses.


\end{abstract}


\vspace{-0.08in}
\section{Introduction}\label{sec:intro}

In the power transmission network, temporary faults may be caused by momentary line contact with vegetation or animals. Resultant faults may have high fault impedance and are cleared without the action of any protective element, making the localization task especially challenging. Frequently recurring disturbances in close proximity to each other might indicate the presence of system vulnerabilities, which may result in catastrophic failures in the future. Therefore, localizing disturbances and rectifying them in time is a critical requirement for safe and reliable grid operations. Traditional methods for disturbance localization has included impedance measurement and travelling wave based approaches which are sensitive to line parameters or have high sampling requirements \cite{review}. The large-scale deployment of Phasor Measurement Units (PMU) in recent years motivates exploring data-driven approaches for localizing transient faults \cite{location, genpmu,Deka}. In \cite{genpmu}, the authors propose locating load increase events using logistic regression. A neural network based approach is also proposed in \cite{Deka} for complementing conventional methods.

\begin{figure}
    \centering
    \includegraphics[width=0.4\textwidth,trim=1.7in 3.6in 1.8in 3.6in,clip]{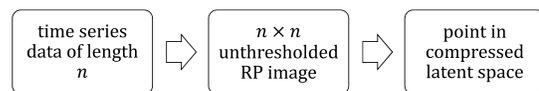}
    \caption{The DeVLearn framework}
    \label{fig:devlearn}
    \vspace{-0.3in}
\end{figure}

In recent years, deep learning (DL) has been gaining acceptance and popularity in the power systems community due to its capability of learning highly non-linear relationships among variables. In recent literature, DL has been used for applications like event classification~\cite{FNET}, dynamic security assessment~\cite{HidalgoArteaga2019DeepLF}, 
and load forecasting~\cite{7778779}. 
This paper proposes a \textbf{De}ep \textbf{V}isual \textbf{Learn}ing framework (DeVLearn) for the transient fault localization task, particularly identifying the faulted line. Once the line is identified, analytical methods, like the one proposed in \cite{location} may be used to  pinpoint the exact fault location.  Our primary objective is to compute a compressed representation of measurement data in a lower dimensional space by training a deep Variational Auto-Encoder (VAE)~\cite{VAE}. In DeVLearn (Fig.~\ref{fig:devlearn}), this is done by first embedding a time series of length $n$ into a $n\times n$ image using unthresholded Recurrence Plots (RP)~\cite{RPOG}. The image is then compressed to a single point in a lower order $k$-dimensional space, also called `latent space'. The image embedding step enables direct application of DL models from the computer vision domain. 

Performance of DeVLearn is demonstrated using measurements from two generator buses of the IEEE 68 bus network. Temporary three-phase faults with different clearing times and fault impedances are simulated on two different lines. Here, we explicitly choose only generator buses as they are more likely to be instrumented with PMUs in reality. As DeVLearn is trained, machine responses to different faults separate into well-defined clusters in latent space. 

The main contributions of the paper are summarized as follows. 1) We show how the  unthresholded RP method may be used to represent univariate time series as images that preserve the temporal relationship in the original time series. 
2) We introduce a DL framework to compute a compressed representation of RP images in low dimensional latent space. We show that the deep VAE is able to separate fault measurements into discernible clusters in this latent space, and hence off-the-shelf classifiers may be used to localize the faults. This dimensionality reduction or feature learning technique is a novel contribution in the power system domain and has immense potential even beyond the fault localization task. 


The remainder of the paper is organized as follows. In section \ref{sec:method} we explain the DeVLearn framework and its components in detail. Section \ref{sec:results} describes our experimental results. Section \ref{sec:conclusion} discusses limitations of the present work, outlines future research directions and concludes the paper.
\vspace{-0.06in}
\begin{figure*}[ht]
    \centering
    \begin{subfigure}{0.32\textwidth}
    \includegraphics[width=0.9\textwidth,trim= 0 1in 0 0,clip]{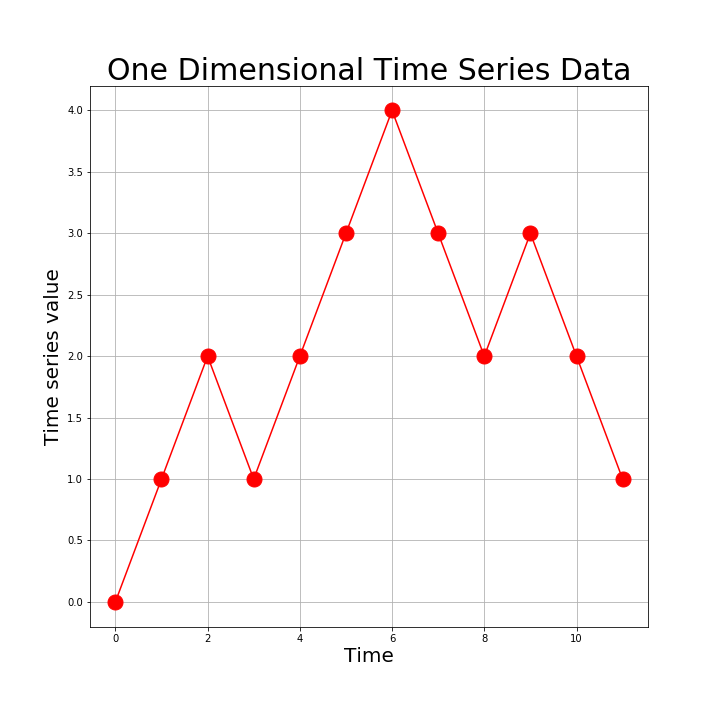}
    \caption{Time series data}
    \end{subfigure}
    \begin{subfigure}{0.32\textwidth}
    \includegraphics[width=0.9\textwidth,trim= 0 1in 0 0,clip]{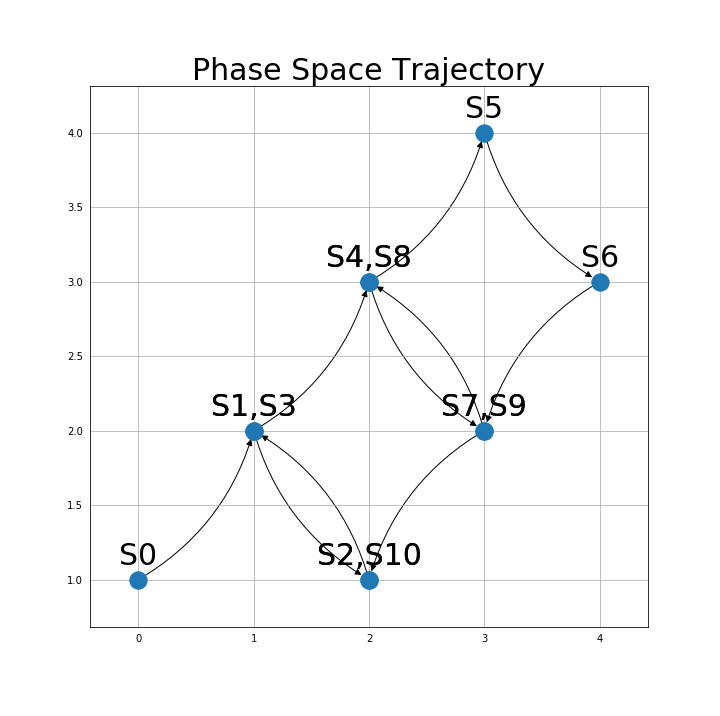}
    \caption{Two dimensionsal phase space trajectory}
    \end{subfigure}
    \begin{subfigure}{0.32\textwidth}
    \includegraphics[width=0.9\textwidth,trim= 0 1in 0 0,clip]{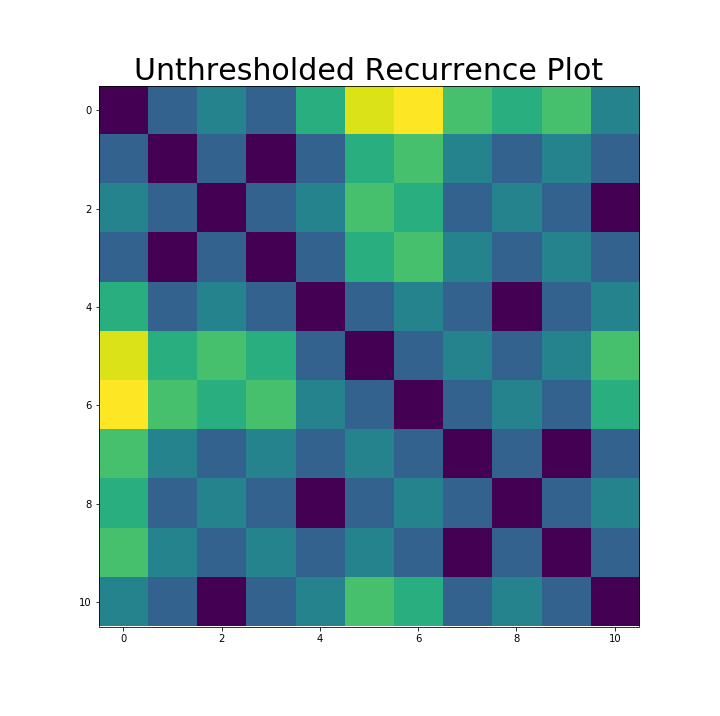}
    \caption{Unthresholded Recurrence plot}
    \end{subfigure}
    \caption{Procedure for constructing unthresholded RP images from time-series data, reproduced from \cite{Hatami2017ClassificationOT}. On the left panel, we show a simple univariate time series $f(t)$ with 12 samples. The middle panel shows its two dimensional phase space trajectory with delay embedding 1. The dots are system states such that $s_i:(f(i),f(i+1))$. The right panel shows the unthresholded RP for $f(t)$. It is a $11\times11$ matrix, whose $(i,j)$-th entry is the euclidean distance between $s_i$ and $s_j$ in the phase space.}  
    \label{fig:RP}
    \vspace{-0.2in}
\end{figure*}

\section{Methodology}\label{sec:method}
Recent years have seen DL achieve major breakthroughs in the fields of computer vision and speech recognition \cite{Goodfellow-et-al-2016,goodfellow2016nips}. DL approaches for time series analysis, however, has been comparatively limited so far. Some deep generative models have been proposed for learning underlying structures in one-dimensional time series data \cite{VDE}, but their performance is heavily dependent on hyperparameter tuning. We propose to leverage image processing advancements  in the power domain by first converting measurements to RP images and then training a DL model to recognize latent structures in them. The efficiency of using RP-based learning for time series classification (TSC) has been demonstrated in \cite{Hatami2017ClassificationOT}. Here, the authors show that RP-embedding is more efficient for TSC than other benchmark methods as well as other image embedding methods proposed in literature \cite{GADF}.

\subsection{Recurrent Plots}\label{sec:RP}
Time series data are characterized by distinct behavior like periodicity, trends and cyclicities. Dynamic nonlinear systems exhibit recurrence of states which may be visualized through RPs. First introduced in \cite{RPOG}, RPs explore the $m$-dimensional phase space trajectory of a system by representing its recurrences in two dimensions. They capture how frequently a system returns to or deviates from its past states. Mathematically, this may be expressed as below.
\begin{equation}
    R_{i,j}=||\Vec{s_i}-\Vec{s_j}||, \quad i,j=1,2,\dots K
\end{equation}
Here, $\Vec{s_i}$ and $\Vec{s_j}$ represent the system states at time instants $i$ and $j$ respectively. $K$ is the number of system states considered. In the original RP method, the $R$ matrix is binary, i.e. its entries are $1$ if the value of $||\Vec{s_i}-\Vec{s_j}||$ is above a pre-determined threshold and $0$ otherwise. We do away with the thresholding since unthresholded RPs capture more information. Images so obtained capture patterns which may not be immediately discernible to the naked eye. A detailed procedure for constructing a RP plot of a simple time series is shown in Fig.~\ref{fig:RP}. 
\begin{figure}[ht]
    \centering
    \includegraphics[width=0.38\textwidth]{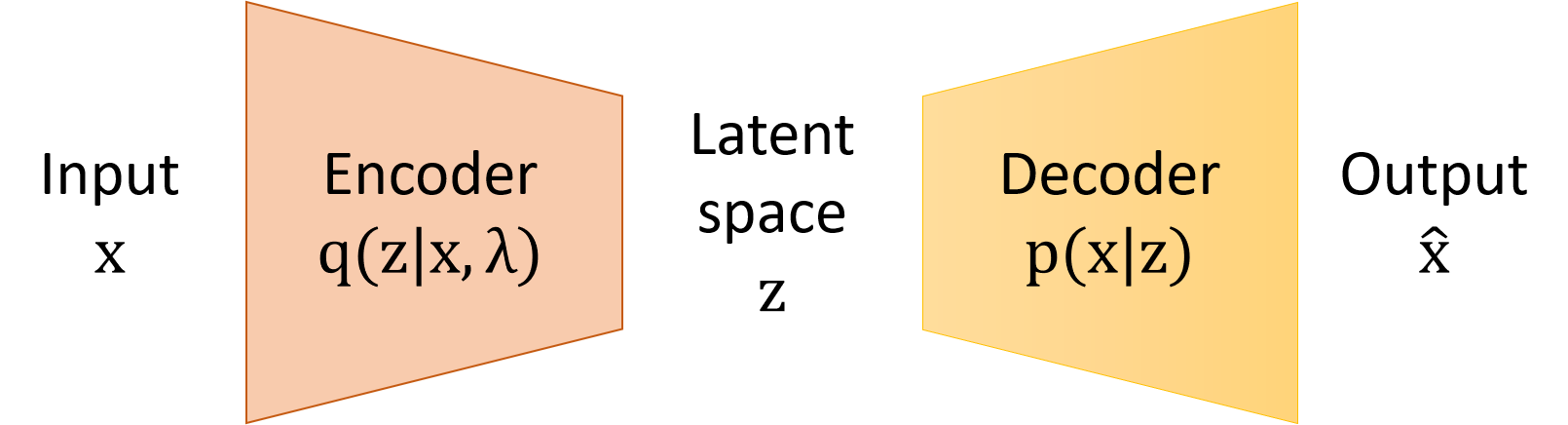}
    \caption{VAE architecture}
    \label{fig:vae-arch}
\end{figure}

\begin{figure}
    \centering
    \includegraphics[width=0.7\columnwidth,trim= 0.5in 3.7in 6.5in 1.5in,clip]{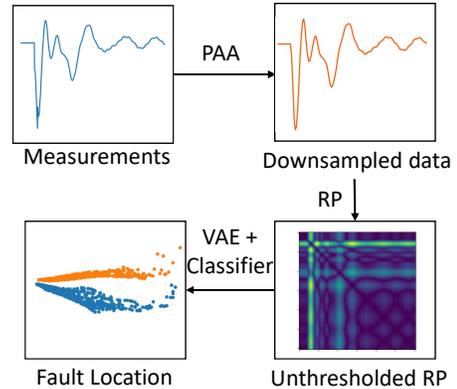}
    \caption{Pipeline showing  DeVLearn operation}
    \label{fig:pipeline}
    \vspace{-0.3in}
\end{figure}

\subsection{Variational Autoencoder}\label{sec:VAE}
An autoencoder (AE) is an unsupervised learning technique where a neural network (NN) is trained to generate outputs that replicate its inputs~\cite{Goodfellow-et-al-2016}. Particularly, a `bottleneck' in the NN architecture is leveraged to create a lower dimension representation of the inputs in the \emph{latent space}. AEs comprise of two components- a) an encoder that learns a compressed representation of the input data (dimensionality reduction), and b) a decoder that learns to reconstruct input data from the compressed representation.

A VAE uses variational inference to generate the distribution of latent variables in the lower dimensional space~\cite{vae_2017}. The distribution of latent variables $z$ for a given input data $x$ follows the posterior distribution $p(z|x)$. Computing $p(z|x)$ in closed form results in an intractable integral. To this end, the variational inference method uses a different distribution $q(z|x,\lambda)$ to approximately infer the computationally intractable distribution $p(z|x)$. This is ensured by training the VAE such that the KL-divergence between the distributions $q(z|x,\lambda)$ and $p(z|x)$ is minimized. It can be shown that the distribution $q(z|x,\lambda)$ which approximates $p(z|x)$ minimizes the expression in (\ref{eq:obj1}).
\begin{equation}
    \arg\min_{q(z|x,\lambda)}-\mathbb{E}_{q(z|x,\lambda)}\log p(x|z)+\mathbb{KL}(q(z|x,\lambda)||p(z))
    \label{eq:obj1}
\end{equation}
The first term in (\ref{eq:obj1}) is the reconstruction loss or expected negative log-likelihood for input data $x$. A small value of reconstruction loss denotes that the VAE is able to accurately reconstruct $\hat{x}$ from input data $x$. The second term is the KL-divergence between the learned distribution $q_{\lambda}(z|x)$ and the prior distribution $p(z)$ which acts as a regularizer term. $\lambda$ is the variational parameter which indexes a family of distributions. In our case, we assume the prior distribution $p(z)$ to be Gaussian which makes the parameter to be mean and variance of latent variables $\lambda=(\mu_{z},\sigma_{z})$ for each input data point $x$.

Just like an AE, a VAE also consists of an encoder, decoder and a loss function. The encoder is a NN which generates parameters $\lambda$ for the distribution $q(z|x,\lambda)$. The decoder is another NN trained to reconstruct the input data $x$ from a given latent representation $z$. The loss function is a weighted sum of reconstruction loss and KL-divergence terms. 
Choosing a weight corresponding to the reconstruction loss which is significantly higher than the other results in overfitting of the VAE, whereas a higher weight for the KL-divergence term enforces the distribution $q(z|x,\lambda)$ to follow the prior distribution $p(z)$.

\subsection{DeVLearn Framework}
The DeVLearn framework puts the components discussed above together to achieve a very powerful latent space representation. The pipeline of the framework is shown in fig. \ref{fig:pipeline}. We reiterate the steps involved, for clarity.

\noindent\textbf{Step 1: } To reduce computation burden, we first downsample time series data using Piecewise Aggregate Approximation (PAA). In this paper, we have downsampled the original signal by a factor of five.

\noindent \textbf{Step 2: } The down-sampled data is converted to unthresholded RP images following the procedure described in section \ref{sec:RP}. We  have used a delay embedding of 1.

\noindent \textbf{Step 3: } A Convolution Neural Network (CNN) based deep VAE model is trained to learn the latent space distribution of the fault data\cite{keras}. The latent space is considered to be two dimensional. The encoder has two hidden layers while the decoder has a single hidden layer. The structure of the deep VAE is similar to the example available in~\cite{kerascode}. 
     
As explained in section \ref{sec:VAE}, the VAE loss function has two elements- reconstruction loss (RL) and KL-divergence term (KLD). The mean squared error (MSE) metric between input data $x$ and reconstructed data $\hat{x}$ has been used for the RL term. Since our downstream application desires well-separated clusters in the latent space, we reduce the relative weight for the KLD term. Potential for other downstream applications also exist. VAEs have been employed to generate synthetic images \cite{Goodfellow-et-al-2016}. Methods of recovering original signals from unthresholded RPs have already been proposed in literature \cite{reconstruction}. Therefore, DeVLearn may be modified to generate realistic synthetic PMU data. This is an exciting research direction that the authors want to pursue in the future.

\noindent\textbf{Step 4: }In the latent space learned in step 3, each signal is compressed to a single point in two-dimensional space. The novelty of DeVLearn is in its capability to learn a latent space where measurements corresponding to different fault locations are automatically separated into disentangled clusters, even when the DL moodel has no explicit knowledge of the data labels. Any classifier like Support Vector Machines (SVM) can now be used to determine location of an unseen fault.

\section {Results and Discussions}\label{sec:results}
\begin{figure}[ht]
    \centering
    \includegraphics[width=0.5\textwidth,trim=2in 1.5in 2in 1.5in, clip]{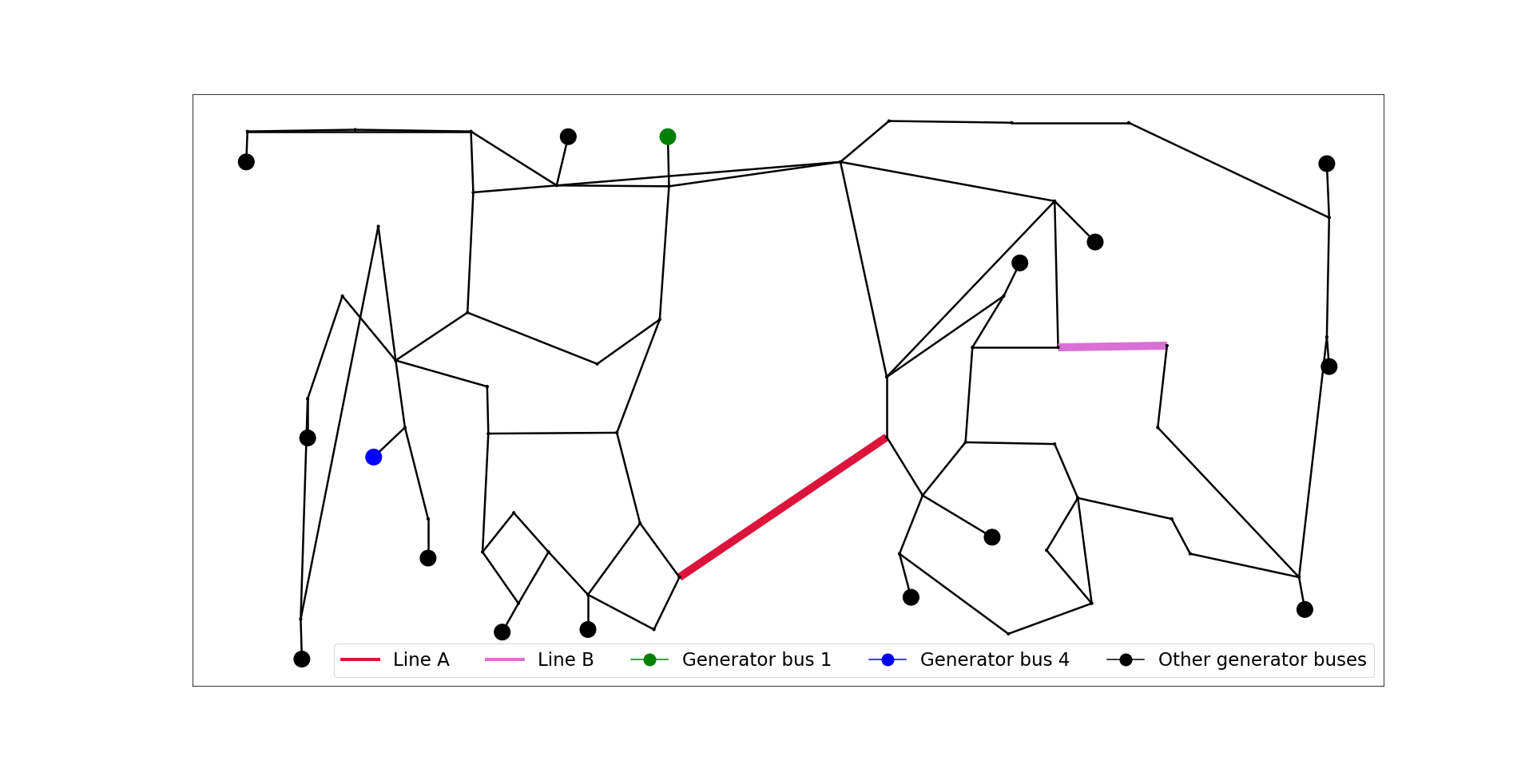}
    \caption{IEEE-68 bus power system with colored edges denoting the two locations of temporary faults. The colored nodes are the generator buses with PMU measurements.}
    \label{fig:network}
\end{figure}
\vspace{-0.2in}

\subsection{Experimental Setup}
In this paper, we analyse transient faults on two transmission lines A and B in the standard IEEE-68 bus power system. To this end, $1000$ three-phase faults are simulated for each line and the voltage magnitudes at two generator buses  (Generator 1 and 4) are recorded. Location of the generator buses and faulted lines is shown in Fig~\ref{fig:network}. The fault impedance for each event is randomly sampled from a uniform distribution between 0 and 1000 Ohms. Similarly, the fault duration is assumed to be uniformly distributed over 10 to 20 cycles of power systems frequency. The simulated events are split into training and testing datasets. Training  and testing datasets have 1800 and 200 events repectively. All power systems simulations are carried out in PSS/E. DeVLearn is trained using the GPU hardware accelaration option available on Google Colaboratory. Training for a batch size of 100 took around 550 $\mu$s for a single epoch.

It must be mentioned here that detecting presence of temporary faults has not been considered in the scope of DeVLearn. Multiple methodologies have been proposed to detect temporary disturbances in literature \cite{jbr2013}.

\subsection{Recurrent Plots for Faults}
\begin{figure}
    \captionsetup[subfigure]{justification=centering}
    \centering
    \includegraphics[width=0.4\textwidth,]{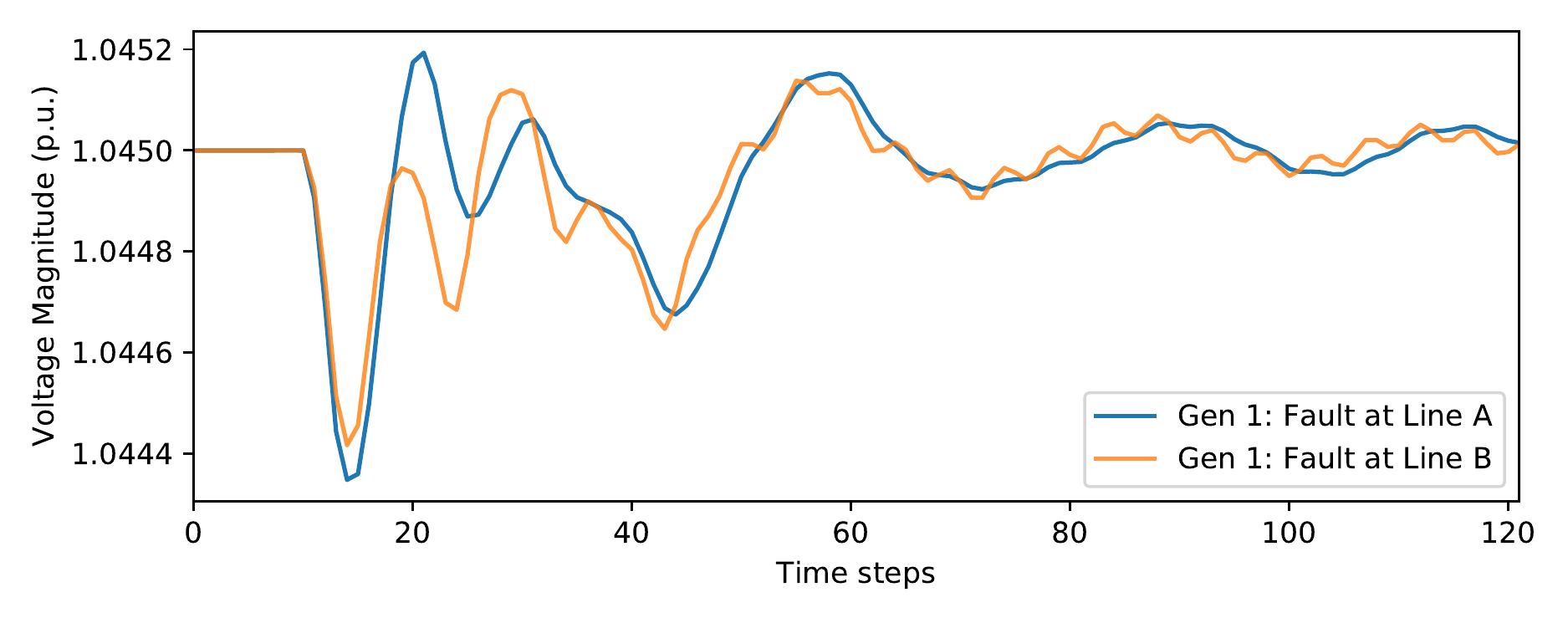}
    \caption{Voltage at Gen. 1 for faults at lines A and B}
    \label{fig:G1}
    \includegraphics[width=0.4\textwidth,]{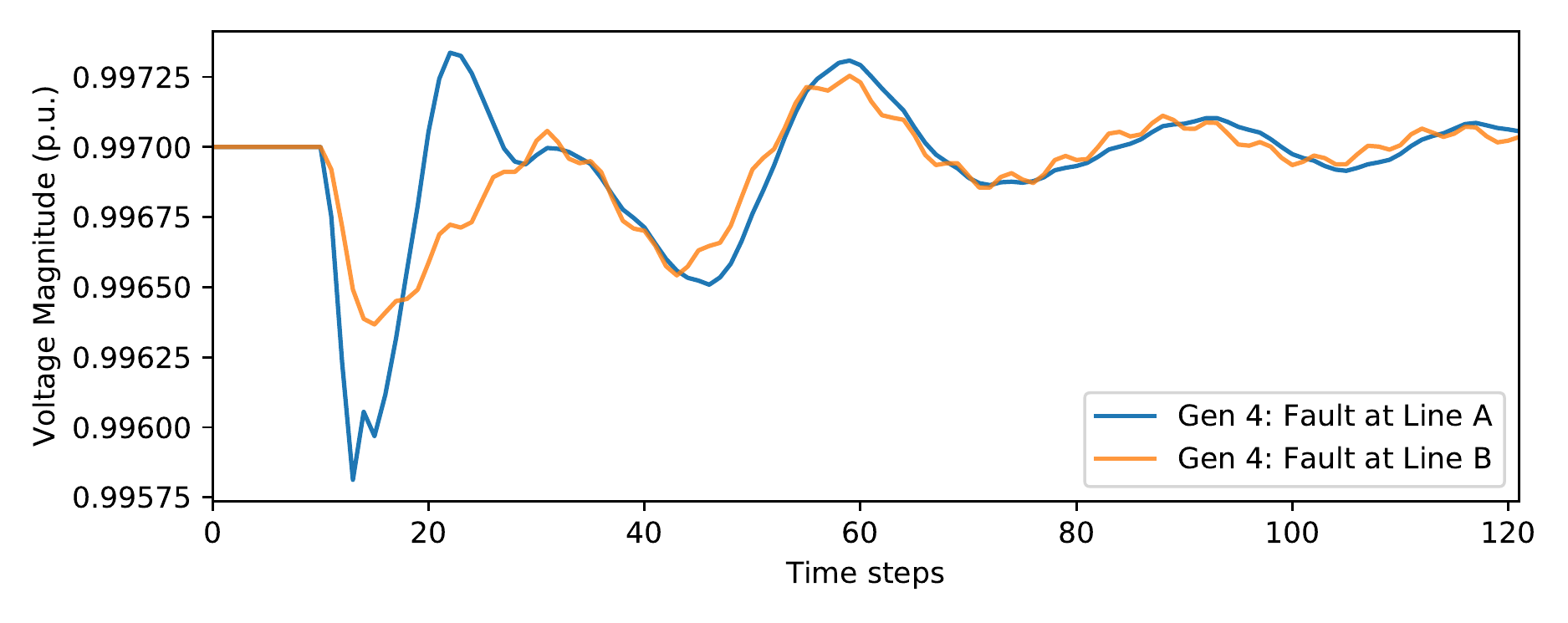}
    \caption{Voltage at Gen. 4 for faults at lines A and B}
    \label{fig:G4}
    \includegraphics[width=0.48\textwidth,]{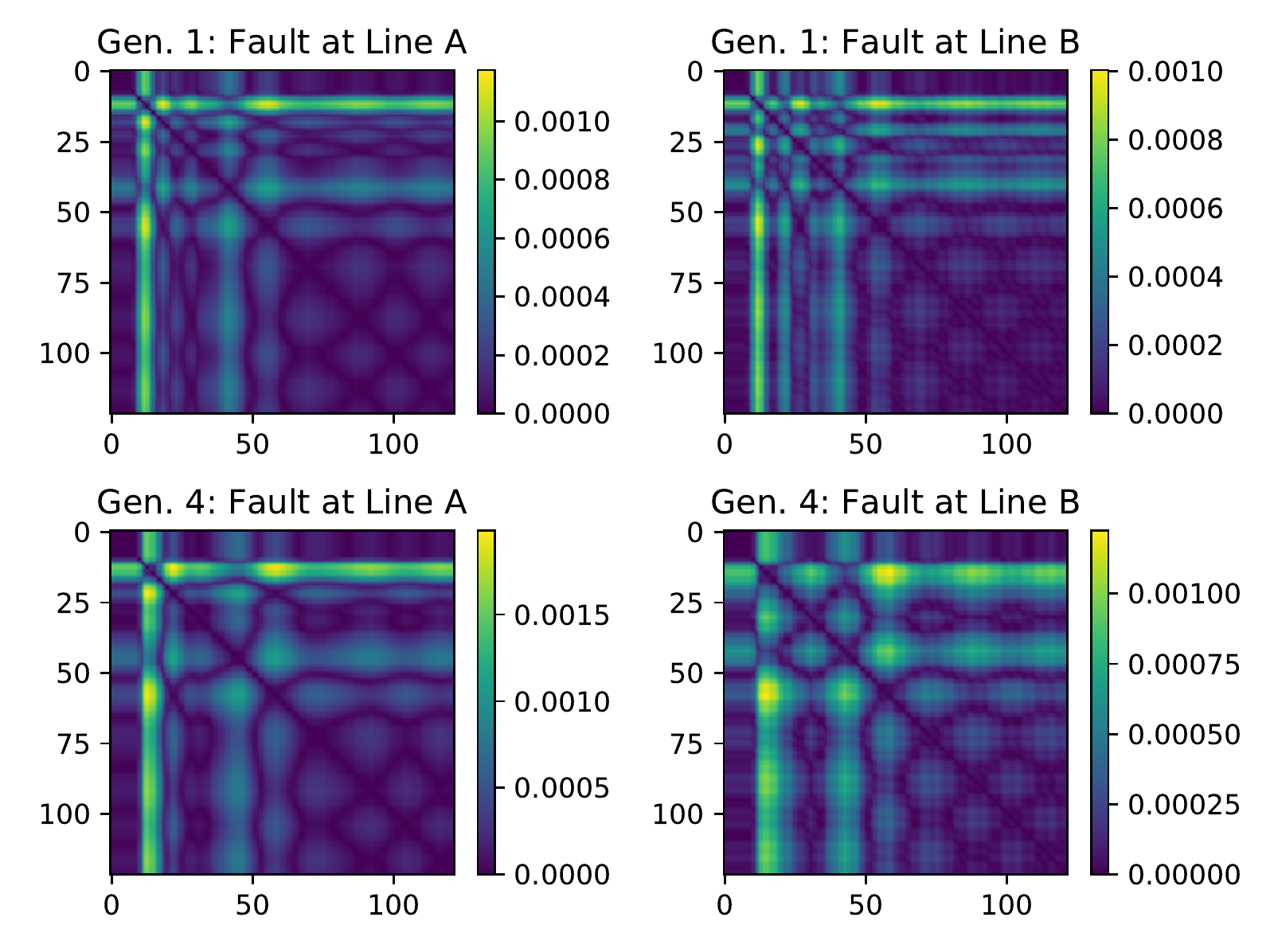}
    \caption{Unthresholded RP images for measurements shown in Fig.~\ref{fig:G1} and Fig.~\ref{fig:G4}.}
    \label{fig:GRP}
    \vspace{-0.2in}
\end{figure}

In order to better understand how generator response to fault events translate to RP images, let us look at the RPs for voltage magnitude measurements at generator buses 1 and 4 for two faults at lines A and B. Fig.~\ref{fig:G1} and~\ref{fig:G4} respectively show the time series measurements (downsampled by a factor of 5) at buses 1 and 4, while Fig.~\ref{fig:GRP} shows the corresponding RP images. In these images, time progresses in a diagonal manner, from the upper left corner to the lower right. It is evident that RP images for the different events may be distinguished, even by the naked eye. Preliminary exploration revealed that images for similar events indeed look similar, even for high impedance faults, where the voltage deviation at generator buses is not very high. The objective now is to teach DeVLearn to recognize the RP images and associate them with the events they correspond to.

\subsection{Training the VAE}
\begin{figure*}[ht]
    \centering
    \begin{subfigure}{0.25\textwidth}
    \includegraphics[width=\textwidth,trim=0.7in 0.7in 0.7in 0.7in, clip]{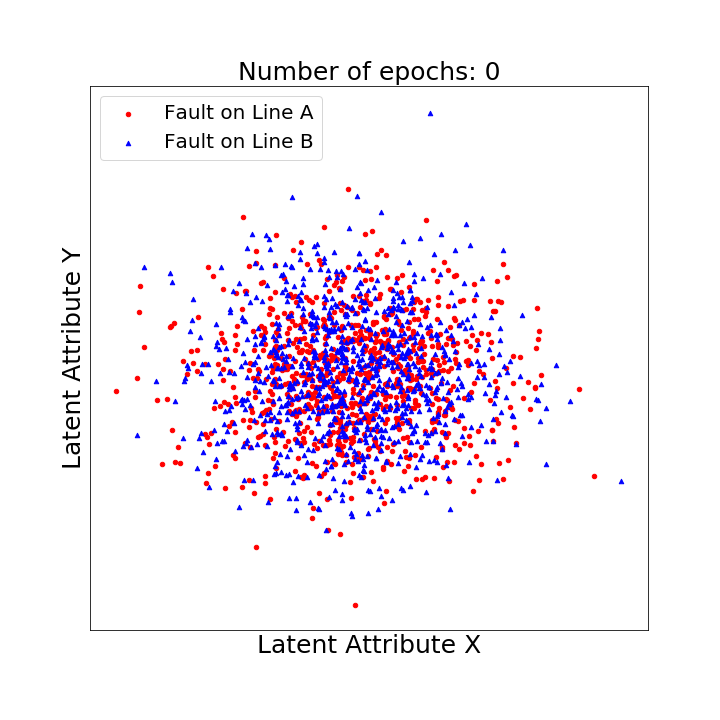}
    \caption{}
    \label{sfig:m1epoch0}
    \end{subfigure}
    \begin{subfigure}{0.25\textwidth}
    \includegraphics[width=\textwidth,trim=0.7in 0.7in 0.7in 0.7in, clip]{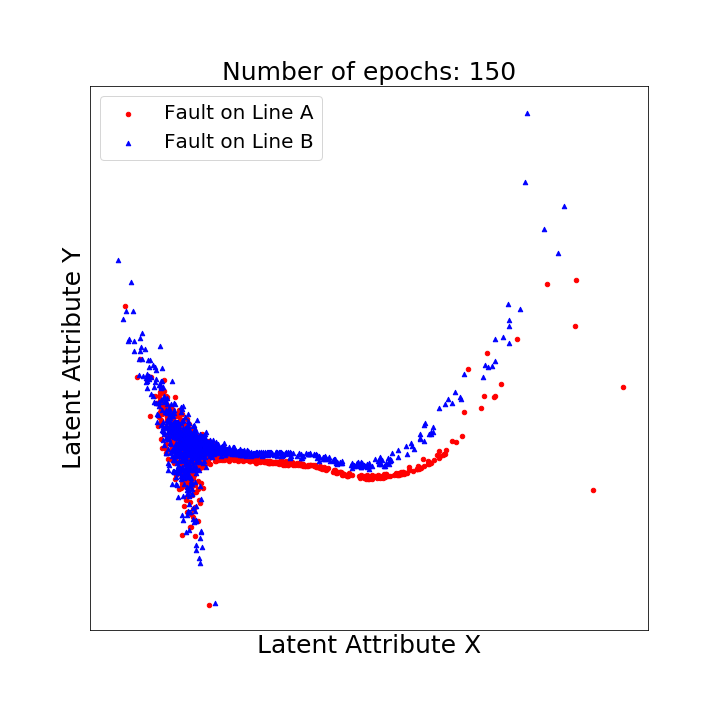}
    \caption{}
    \label{sfig:m1epoch150}
    \end{subfigure}
    \begin{subfigure}{0.25\textwidth}
    \includegraphics[width=\textwidth,trim=0.7in 0.7in 0.7in 0.7in, clip]{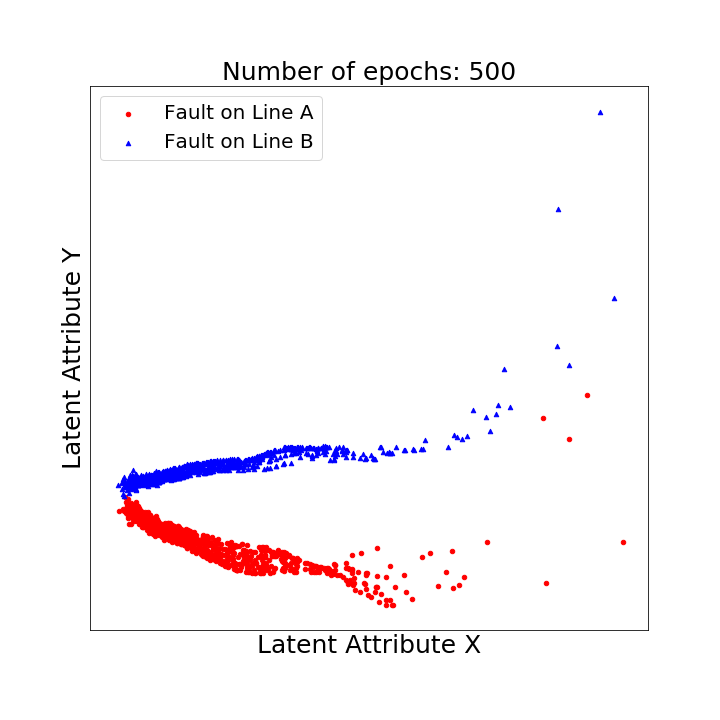}
    \caption{}
    \label{sfig:m1epoch500}
    \end{subfigure}
    \begin{subfigure}{0.25\textwidth}
    \includegraphics[width=\textwidth,trim=0.7in 0.7in 0.7in 0.7in, clip]{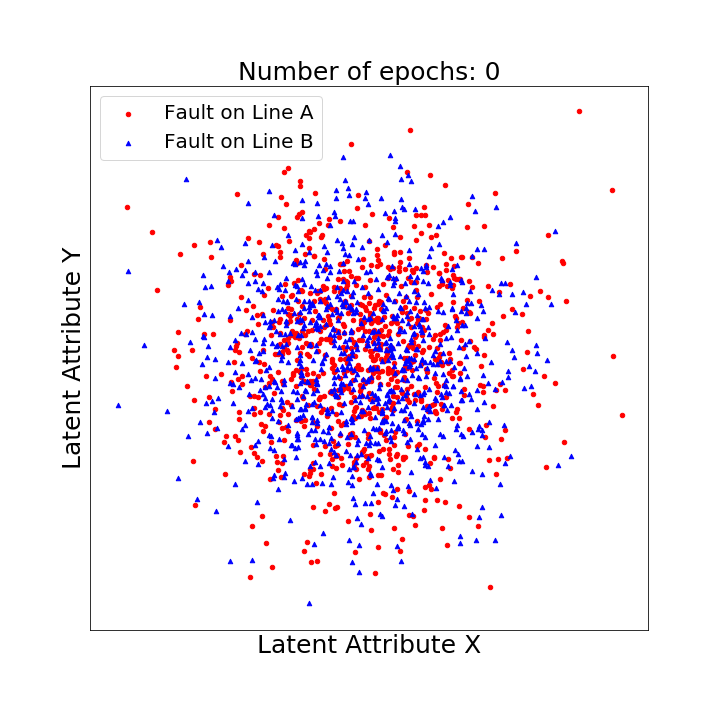}
    \caption{}
    \label{sfig:m4epoch0}
    \end{subfigure}
    \begin{subfigure}{0.25\textwidth}
    \includegraphics[width=\textwidth,trim=0.7in 0.7in 0.7in 0.7in, clip]{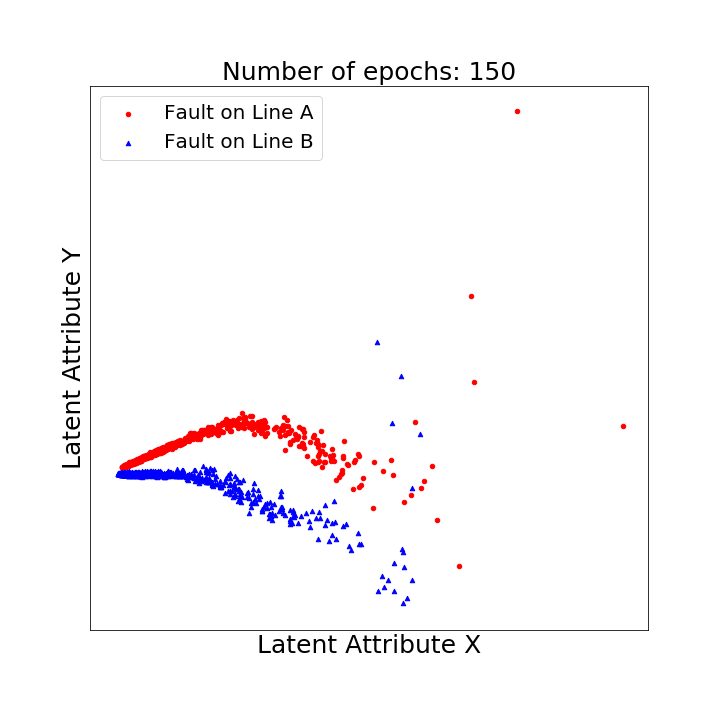}
    \caption{}
    \label{sfig:m4epoch150}
    \end{subfigure}
    \begin{subfigure}{0.25\textwidth}
    \includegraphics[width=\textwidth,trim=0.7in 0.7in 0.7in 0.7in, clip]{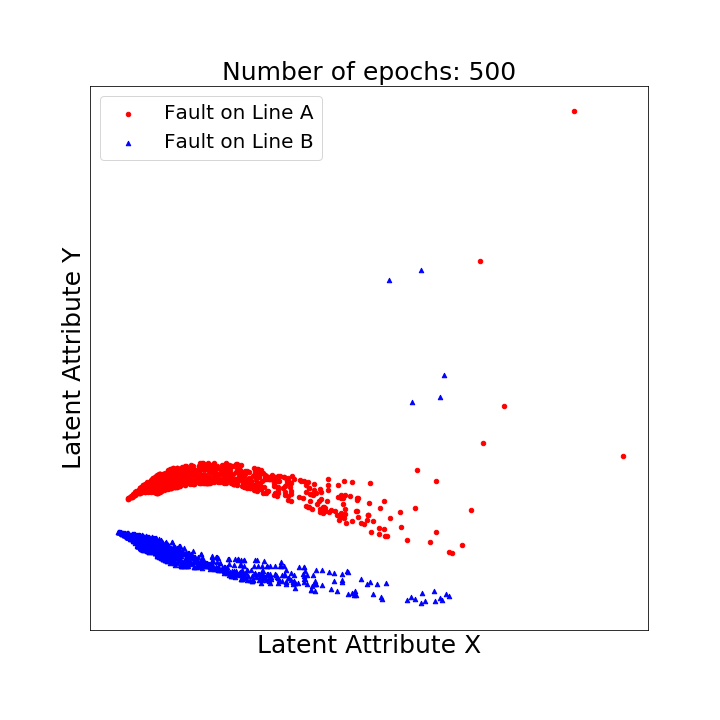}
    \caption{}
    \label{sfig:m4epoch500}
    \end{subfigure}
    \caption{Evolution of latent attribute distribution in the compressed two dimensional space over different training epochs. Figs.~\ref{sfig:m1epoch0}-\ref{sfig:m1epoch500} shows the evolution of the latent space when the DeVLearn framework is trained with voltage magnitude measurements at generator bus 1. The distribution of latent space attributes for measurements at generator bus 4 is shown in Figs.~\ref{sfig:m4epoch0}-\ref{sfig:m4epoch500}. It can be seen that as training epoch progress, measurements corresponding to faults at different locations separate into discernible clusters in the latent space.}
    \label{fig:epochs}
    \vspace{-0.3in}
\end{figure*}

The deep VAE component of DeVLearn is trained using 1800 instances of $120\times120$ grayscale images. Each training epoch uses a batch size of 100 data points. A separate DeVLearn framework is trained for each of the generators, but the VAE architecture and loss functions were not altered. The encoder projects the training set into a compressed two dimensional latent space, whose evolution with training epochs is shown in Fig.~\ref{fig:epochs}.  It can be clearly seen that data from different faults start separating into clusters as training progresses, and significant separation is achieved at 500 epochs from both the models. Therefore, we are able to form an estimate of fault location using only local voltage magnitude measurements.

\subsection{Determining Fault Location}
\begin{figure}
    \captionsetup[subfigure]{justification=centering}
    \centering
    \begin{subfigure}{0.24\textwidth}
    \centering
    \includegraphics[width=0.9\textwidth]{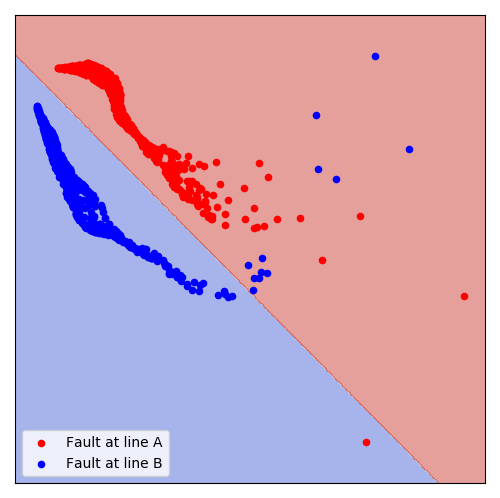}
    \caption{Classifying measurements from Gen. bus 1: training data}
    \label{sfig:g1svctrain}
    \end{subfigure}
    \begin{subfigure}{0.24\textwidth}
    \centering
    \includegraphics[width=0.9\textwidth]{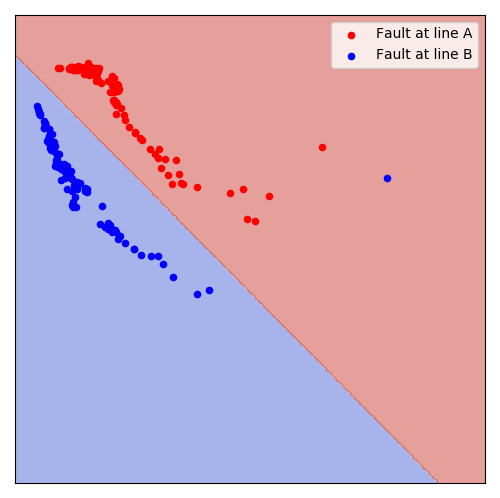}
    \caption{Classifying measurements from Gen. bus 1: testing data}
    \label{sfig:g1svctest}
    \end{subfigure}
    
    \begin{subfigure}{0.24\textwidth}
    \centering
    \includegraphics[width=0.9\textwidth]{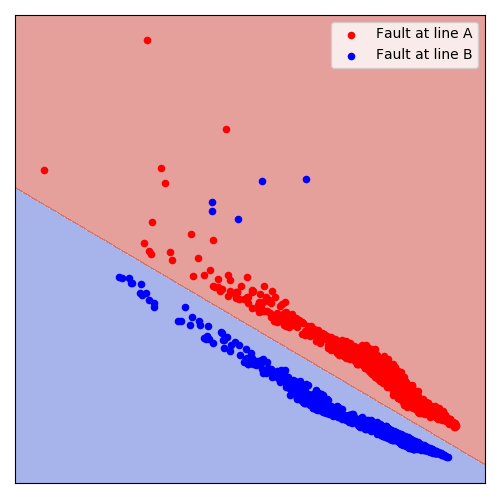}
    \caption{Classifying measurements from Gen. bus 4: training data}
    \label{sfig:m4train}
    \end{subfigure}
    \begin{subfigure}{0.24\textwidth}
    \centering
    \includegraphics[width=0.9\textwidth]{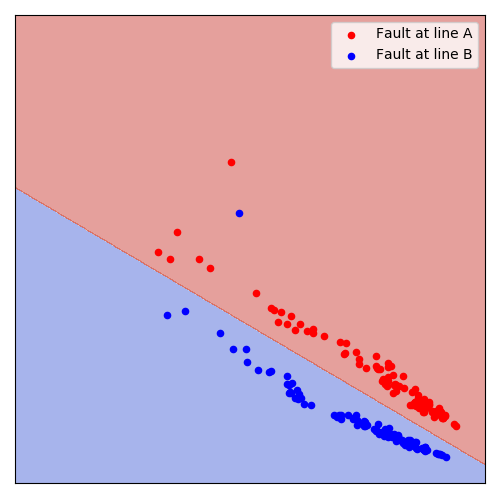}
    \caption{Classifying measurements from Gen. bus 4: testing data}
    \label{sfig:m4test}
    \end{subfigure}
    \caption{Using a SVM classifier with linear kernel to classify latent space learned by DeVLearn}
    \label{fig:svc}
    \end{figure}

We check the performance of a SVM classifier with linear kernel on the latent space learnt by the DeVLearn framework after 1000 training epochs. The resultant decision boundary is shown in Fig.~\ref{fig:svc}. It is evident that in the latent space, fault data for two lines are almost linearly separable. With a linear SVM classifier, we obtain a training accuracy of 99.33\%  and 99.72\% for generator 1 and 4 respectively. Testing accuracy for both generators is 99.5\%. Although a classifier with a non-linear kernel (Radial Basis Function or RBF kernel, for instance) would have achieved higher accuracy, we intend to show that sophisticated classifiers are not required to achieve good performance.

\vspace{-0.08in}\section{Conclusion}\label{sec:conclusion}

This paper provides a proof of concept that image embedding aided deep learning may be used to determine temporary fault locations in power systems with high accuracy. We demonstrate the capability of the proposed framework DeVLearn in learning useful information from unlabeled univariate time series data in the context of distinguishing faults at different locations. This is lucrative, keeping in mind the limited availability of labeled data in the power domain. Research scope exists in devising image embedding strategies for multivariate time series data. Of course, more tests with faults at different lines, network topologies and operating conditions are required to place higher confidence in DeVLearn, and this is a direction that the authors are pursuing. The idea is to validate the DeVLearn framework with actual PMU data and expand it to applications beyond fault localizing, for example generating realistic synthetic PMU data.


\bibliographystyle{./bibliography/IEEEtran}
\bibliography{./bibliography/IEEEabrv,./bibliography/IEEEexample}

\end{document}